# Observation of domain wall motion in a polycrystalline vortex lattice

Malcolm Durkin, Ian Mondragon-Shem, Taylor L. Hughes, Nadya Mason

*We present measurements showing how disorder determines the dynamics of a vortex lattice. Using superconductor-normal-superconductor (SNS) arrays placed in finite magnetic fields, disorder is introduced by shifting the field away from values where the vortex lattice is commensurate with the SNS array. By applying a current to drive vortex motion, we observe a two-step transition at incommensurate fields: first from pinned vortices to lattice defect motion, and then from individual defect to bulk vortex lattice flow. This behavior is consistent with a polycrystalline vortex structure where defects form on the edges of crystalline domains. This demonstrates that a disordered, interacting vortex system can favor interface physics rather than a quasiperiodic regime.*

How crystalline order is destroyed in the presence of disorder is a fundamental question in condensed matter physics and determines the structure of a wide range of systems, including solids and magnetic materials. Disorder typically disrupts crystals in two ways: by creating interface structures like domain walls in polycrystalline materials [1, 2] or by developing new bulk structures like disordered glasses [3]. Vortex systems provide excellent test-beds for studying the introduction of disorder into crystalline structures. A perfectly ordered type-II superconducting film has a crystalline Abrikosov lattice [4,5], but disorder in the film can destroy this crystalline structure and result in unusual phases, as demonstrated in the much-studied temperature driven transition from a quasiperiodic "Bragg glass" [3,6] into a polycrystalline state [7, 8,9,10].

However, it has been a challenge to controllably introduce disorder into superconducting films. One controlled way of studying disorder is to consider systems of particles in a periodic potential, where disorder is introduced when the particle lattice does not match the potential well. This type of system is described by the 2D Frenkel-Kontorova (FK) model, which exhibits a rich set of structures and modes of motion. Molecular dynamics simulations predict motion dominated by lattice defects known as kinks; these include point defect motion [11], domain wall or grain boundary motion [12], and quasiperiodic compression zone motion [13,14] depending on the structure of the arrays. Experimentally, the motion of point defects has been studied in artificial pinning center arrays (APCs) on superconducting films, which show signatures of both linear and turbulent defect motion [15,16,17], and in optically trapped colloid lattices, which have demonstrated the motion of quasiperiodic compression zones [18].

In this paper, we use superconductor-normal-superconductor (SNS) arrays as an experimental realization of the FK model and examine transitions between ordered and disordered states. SNS arrays provide vortices with a periodic potential defined by array geometry [19] and allow the vortex filling fraction of each potential well to be controlled using magnetic field [20]. This system differs from APC arrays, which represent the narrow potential well limit of the FK model, and are typically studied when the vortices greatly outnumber the pinning sites; its behavior more closely resembles optically trapped colloid lattices. Despite advantages in rapid imaging, colloid lattice studies usually occur at a limited number of fillings and with coarse transport measurements. In contrast, we study vortex transport in SNS arrays at incrementally different fillings and are able to observe the transition from commensurate to incommensurate fillings. Previous SNS array studies have focused crystalline vortex array structures, which form when the vortex lattice is commensurate with the array's potential wells and can be identified via electrical transport measurements as dips in magnetoresistance [20] and peaks in de-pinning currents [21]. The structure of the vortex lattice and type of collective vortex motion at incommensurate fillings is less well studied. Here, we show through transport measurements that SNS arrays demonstrate two-step vortex transitions at incommensurate fillings: first from pinned to lattice defect motion, then from lattice defect to bulk lattice motion. Comparing our measurements with a dynamic molecular vortex model, we show that this two-step transition is indicative of domain wall motion in a polycrystalline vortex lattice and not a vortex glass. We thus demonstrate that a disordered, interacting vortex system with a sufficiently strong periodic pinning potential can favor interface physics rather than a quasiperiodic regime, as suggested by Ref [12].

The devices we measure consist of triangular arrays of superconducting Nb islands on normal metal films. The normal metal film (1nm Ti, 10 nm Au) is patterned in a four-point measurement configuration. A triangular array of Nb islands that are 70 nm thick, 260 nm in diameter, and have 490 nm edge-to-edge spacing is then patterned on top of the normal metal film using electron beam lithography and electron beam evaporation. Transport measurements of these arrays show that the entire array becomes superconducting below a transition temperature $T_c$ [22] with a zero-field transition that can be associated with a Berezinski-Kosterless-Thouless transition [23]. The superconducting transition is shown in Fig. 1 (a). Measurements in this paper are performed at temperatures well below $T_c$, where the array exhibits magnetoresistance oscillations at finite fields consistent with vortex transport [24].

Samples are studied using four-point measurements [shown in Fig. 1(a) inset] in a dilution refrigerator at 17 mK, sweeping DC current and measuring voltage. Vortices experience a periodic potential from the island array [19], where the local energy minimum is at the center of each triangle formed between adjacent islands, and an energy barrier exists at the array edges. The current applies a Lorentz force on the vortices. If sufficient to overcome the energy barrier, the Lorenz force will de-pin vortices and drive their motion, which is measured as a voltage across the sample. The number of vortices per triangle is determined by the magnetic field and island spacing, and is characterized by the number of flux quantum per plaquette (the "filling fraction"), or $f=\Phi/\Phi_0$, where $\Phi$ is the flux through a unit cell and $\Phi_0 = h/2e$ is the quantum of flux.

The dilute filling behavior of the arrays (i.e., at small $f$ values) is shown in Fig. 1(b), a plot of $dV/dI$ as a function of $f$ for different $I$ (obtained by taking the derivative of an $I$-$V$ measurement). In this regime, the array undergoes a current driven transition from pinned vortices ($dV/dI=0$) to flux flow (constant $dV/dI$) at a de-pinning current [24]. The flux flow—or lattice flow—regime occurs when all vortices are moving at a terminal velocity, where the Lorentz force is equal to a dissipative force, resulting in a linear relationship between $I$ and $V$. This leads to a constant $dV/dI$ that we refer to as the flux flow resistance, $R_{ff}$. $R_{ff}$ scales linearly with filling fraction $f$, as the measured $V$ is proportional to the total number of moving vortices.

This is shown in Fig. 1(b), where, for sufficient currents and low $f$ values, the d$V$/d$I$ curves fit to a single black line representing $R_{ff}$. When plotted on a broader range in Fig. 1(c), the extrapolated fit to $R_{ff}$ represents an upper limit to the $dV/dI$ measured for a given $f$ and indicates when all vortices are flowing.

At larger $f$ values, $dV/dI$ no longer has a linear relationship with $f$ for most applied currents. This is visible in Fig. 1(b), when a number of curves start to diverge from the $R_{ff}$ fit at $f > 0.05$; the effect is even more pronounced in Fig. 1 (c), where a pattern of peaks and dips emerge when $f > 0.1$. The departure from linear $dV/dI$ occurs because greater vortex density results in stronger vortex-vortex interactions. Vortex-vortex repulsion usually leads to weaker pinning as filling is increased. However, at special fillings—e.g., $f = 1/12, 1/6, 1/4, 1/2$—the vortex lattice is commensurate with the array potential wells, resulting in crystalline vortex orderings and strong pinning, as evidenced by dips in $dV/dI$ and a greater de-pinning current at commensurate fillings.

Commensurately and incommensurately filled lattices exhibit different dynamic behavior. The dynamics as a function of current can be seen in Fig. 1(d), which plots d$V$/d$I$ vs $I$ for the commensurate filling $f=0.25$ alongside the incommensurate pinning $f=0.20$. Driven from pinned to flux flow, the commensurate filling undergoes a single transition, while the incommensurate filling undergoes two distinct transitions separated by an intermediate region of constant $dV/dI$. Similar behavior can be seen in Fig. 1(b), where incommensurate fillings have intermediate clusterings of lines, as exemplified at $f = 0.20$. In contrast the commensurate fillings transition rapidly into flux flow, resulting in peak reversals where low current dips turn into peaks at higher currents (e.g., as at $f = 0.5$). Similar peak reversals have been observed in previous works [7] and used as evidence for a vortex Mott-insulator to metal transition [25].

To characterize incommensurate lattice dynamics, we plot the transition locations as a function of field and current in Fig. 2. Since the transitions are associated with steps in d$V$/d$I$, they can be identified as $d^2V/dI^2$ peaks, or the bright regions of Fig. 2. For commensurate fillings, depinning transitions occur at higher currents, as indicated by the arrows in Fig. 2. While dilute fillings and commensurate fillings only have a single transition (only one visible $d^2V/dI^2$ peak for a given $f$), incommensurate fillings often undergo a two-step transition with two visible $d^2V/dI^2$ peaks. This is evident in Fig. 2, where the incommensurate fillings indicated by dashed lines have first transitions marked by blue circles and secondary transitions marked by red Xs.

The single transitions smoothly split into two transitions as filling is shifted away from commensurate values. This can be seen in Fig. 2, where the single transition at $f=1/6$ smoothly splits into two diverging transition curves as the filling is increased, with the upper curve marked with an "X" and the lower curve marked with an "O" at $f\sim0.21$. Another prominent splitting is also visible as the field is increased from $f=1/4$ and a less prominent secondary transition curve splits off as the filling is decreased from $f=1/6$. The smooth splitting as a function of filling indicates that this behavior is determined by lattice structure, with the vortex lattice transitioning from pinned regime to an intermediate vortex motion regime to a lattice flow regime. This splitting has not been previously discussed and presents implications regarding vortex structure and motion.

To better understand the dynamic behavior in the different regimes, including the $d^2V/dI^2$ peak splitting, we use a molecular vortex model based on the Langevin equation for mutually repulsive vortices in a periodic potential [24], which is a re. This has equations of motion given by

$$m\ddot{x}_i(t) = F_{applied} - \frac{\partial V(x_i(t))}{\partial x_i} - \eta \dot{x}_i(\tau) + \varepsilon_i(t) + \sum_{j=1}^{N} U\left(\frac{x_i(t) - x_j(t)}{L_{int}}\right),$$

where $m$ is a mass term related to capacitance, $F_{applied}$ is the Lorentz force (proportional to the applied current), $V(x_i(t))$ is a periodic potential defined by the array, $\varepsilon_i(t)$ is a stochastic force used to simulate finite temperature, and $U(x)$ is the mutual repulsion between vortices. Low resistance systems such as this one are in the overdamped limit, with $m = 0$ [24]. Further details are found in the supplement. By varying the number of vortices per potential well and applying a driving force, we simulate current sweeps at magnetic field intervals. This model has been previously used to describe the behavior of SNS arrays. Although some features, such as $dV/dI$ peaks at the de-pinning current, are not described by the model, we previously addressed this by including the effects of a history dependent dissipative force [24] (for simplicity, this effect is not included here, but is described in the supplement).

A basic one dimensional (1D) simulation, effectively the 1D FK model, can replicate the two-step transition for incommensurate fillings. Shown in Fig. 3(a), the lattice undergoes a direct transition from pinned to flux flow at commensurate fillings, such as $f$=0.50. Due to the ordered arrangement of vortices [Fig. 3(a) lower right inset], the vortex lattice moves in unison and exhibits dynamics similar to that of a single vortex. As additional vortices are added, defects enter the lattice [Fig. 3(a) upper left inset], separating ordered domains. These defects require a lower depinning force than the rest of the lattice and begin moving prior to the lattice depinning current, resulting in a transition from pinned to defect flow to lattice flow. For a more direct comparison with data, the predicted d$V$/d$I$ is provided in Fig. 3 (b) for the different transport regimes. Notably, the distinct two-step transition occurs when well-defined defects separate ordered regions, analogous to domain walls. In contrast, higher values of disorder than shown in Fig. 3 yield an amorphous state, exhibiting only a single transition [For a detailed analysis, see Supplement].

To simulate a 2D system, we use a periodic potential similar to the one produced by the triangular island array, incrementally sweep current at different vortex populations, and extract the vortex motion. As shown in Fig. 4(a), this simulation reproduces the basic features of the data in Fig. 2: de-pinning current peaks at commensurate fillings and a second d$^2V$/d$I^2$ peak that splits as the filling is altered from commensurate values. Similar to our measurements, the two-step transition shows clearly above $f$=1/6, with the first transition marked with a white circle and a second transitions marked by a white X. The lattice structure is clarified by Fig. 4(b), which shows simulated vortex positions and dynamics in the domain wall motion regime (marked by a black circle in Fig. 4(a)). The black circles and red circles show initial and final vortex locations, respectively, over the short time roughly corresponding to a vortex crossing from one well to another [for a plot of lattice flow, which is above the second transition, see supplement]. Vortex motion occurs primarily in defects, which appear as cracks that form between crystalline structures. Motion along these cracks is visible as the line following the dark circles in Fig. 4(b). The presence of defect lines not only indicates that the intermediate transport regime involves defect motion, but is consistent with a polycrystalline structure where defects appear on the interfaces separating crystalline domains. These results suggest the intermediate transport regime involves domain wall motion.

Thus, the two-step transition we observe at incommensurate fillings is consistent with a transition from pinned vortices to lattice defect motion to lattice flow. Molecular vortex model simulations suggest that this motion occurs on the edge of crystalline domains, providing evidence for domain wall motion and polycrystalline structure in this system similar to those recently predicted in artificial pinning array structures [12,26], but not previously explored in experimental studies.

# Supplemental Materials

### 1. Simulating Vortex-Vortex Repulsion

The magnitude of the vortex-vortex repulsion force is given by $U(X)=K_1(X/L_{int})$, where $K_1(x)$ a modified Bessel function of the second kind. In the long interaction limit ($L_{interaction}>>a$, where $a$ is center to center island spacing), which is applicable for arrays, this approaches $U(X)=C/X$ where $C$ is a constant modifying the magnitude of the repulsion. In order to have the calculations scale in a reasonably efficient way, interactions between vortices over $10a$ away are ignored. To avoid artifacts from vortices entering and leaving interaction ranges of other vortices, we smoothly tapper the interaction range using the following form:

$$U(x) = BU_0(x) = B\left( \frac{a}{x\left[1+\exp\left(\frac{|x|-8a}{2a}\right)\right]} - \frac{1}{10(1+e)} \right) \quad (S1)$$

As shown in figure S1, this is a reasonable approximation for nearest neighbor and next nearest neighbor interactions in the regime of interest, while effectively ignoring interactions further than that.

### 2. One Dimensional Simulation

The one dimensional system is given by the potential,

$$V(x) = \frac{Aa}{2\pi}\sin\left(\frac{2\pi x}{a}\right) - F_I x \quad (S2)$$

Where $A$ controls barrier height and $F_I$ is the force resulting from the applied current. The system being modeled has low resistance and is in the overdamped limit, allowing $m$ to be set to $0$. The differential equation for an overdamped vortex array is then solved using Euler's method. This is an iterative method that repeats the following calculation over short periods of time, $\Delta t$,

$$\dot{x}_i(t+\Delta t) = \frac{1}{\eta}\left( F_I - A\cos\left(\frac{2\pi x_i(t)}{a}\right) + \epsilon_i(t) + B\sum_{j=1}^{N} U_0\left(x_i(t) - x_j(t)\right) \right). \quad (S3)$$

$$x_i(t+\Delta t) = x_i(t) + \dot{x}_i(t)\Delta t. \quad (S4)$$

The free parameter of in this equation is the ratio between the periodic potential parameter, *A*, and the repulsion parameter *B*, a ratio that determines the stiffness of the lattice. The stochastic force, $\epsilon_i(t)$, is obtained using a random number generator with an exponential distribution of the form

$$\exp\left(-\frac{|\epsilon_i(t)|}{kT}\right). \quad (S5)$$

The simulation seeds N vortices in 50 wells (f=N/50) and then performs a slow anneal from high temperature to low temperature to find a ground state configuration, imposing periodic boundary conditions. We then run the simulation starting with this ground state at varying currents. The results can be seen in Figure S2 with two lattice stiffness parameters: B/A=2 and B/A=6. A stiffness of B/A=2 results in broad two step regions around f=0.5 and f=1.0 (dark and light blue). In contrast, a stiffness of B/A=6 has a distinct two step transition only in relatively narrow range around f=0.5 and f=1.0.

The relationship between vortex lattice structure and the presence (or absence) of a two-step transition is investigated in Figure S3, which shows the vortex structure at both B/A=2 and B/A=6. At the f values given, B/A=2 exhibits a prominent two step transition and B/A=6 either exhibits a less visible two step transition or has only a single step, providing a comparison between the two observed transport phenomena. *B/A=2* [Figure S3 (a)(c)] results in well-defined defects that are limited in area to one or two wells: *f=0.54* yields defects in the form of vortices in adjacent wells and *f=0.9* yields defects in the form of empty wells in an otherwise filled lattice. In both cases, the defects can be interpreted as domain walls separating ordered domains, with f=0.54 defects each separating two half-filled domains and the f=0.9 defects each separating two entirely filled lattices.

In contrast, *B/A=6* defects are more difficult to identify spatially. Rather than appear as a pair of adjacent vortices or as an empty well, these defects are groupings of perturbed vortices that no longer rest in the center of the wells. As seen in Figure S3 (b)(d), the defects occur over a region 10 wells wide, with vortices either perturbed towards the center of the defect (*f=0.54*) or away from the center of the defect (*f=0.90*). At filling *f=0.54*, there are some segments of vortices that are unperturbed, allowing for a visible intermediate step. At f=0.9, the defects are close enough for the lattice to take on a quasiperiodic structure and there is only a single transition. Thus, the two step transition is a signature of distinct defects, which form on the interfaces between ordered domains.

### 3. Absence of a differential resistance peak in experimental data

While a differential resistance peak is predicted in the simulations, it is absent in experimental simulations. We have previously addressed the and the associated I-V behavior using a history dependent dissipative force The inclusion of a history dependent dissipative force for a 1D system with B/A=2 is shown in figure S4. Here, the inclusion of the term removes the peaks on both the defect motion and the lattice motion steps[24]. Since this does not fundamentally alter the types of vortex motion occurring, we do not include this term in any other section of this work for simplicity and to save computational resources.

### 4. Two Dimensional Simulation

We create a potential with triangular barriers as well as exclusion zones where the superconducting islands would be using the form

$$V_1(x,y) = \exp\left(-\frac{\left(\mod\left(\frac{x}{a}-\frac{1}{4},1\right)-\frac{1}{2}\right)^2 + \left(\mod\left(\frac{y}{a}+\frac{\sqrt{3}}{4},\sqrt{3}\right)-\frac{\sqrt{3}}{2}\right)^2}{2\sigma^2}\right) + \exp\left(-\frac{\left(\mod\left(\frac{x}{a}+\frac{1}{4},1\right)-\frac{1}{2}\right)^2 + \left(\mod\left(\frac{y}{a}-\frac{\sqrt{3}}{4},\sqrt{3}\right)-\frac{\sqrt{3}}{2}\right)^2}{2\sigma^2}\right) \quad (S6)$$

$$V_2(x,y) = -\left|\sin\left(\frac{4\pi y}{a\sqrt{3}}\right) + \sin\left(\frac{2\pi}{a}\left(x+\frac{y}{\sqrt{3}}\right)\right) + \sin\left(\frac{2\pi}{a}\left(x-\frac{y}{\sqrt{3}}\right)\right)\right|^{N_{shape}} \quad (S7)$$

$$V(x) = A(C_1 V_1(x,y) + C_2 V_2(x,y)) - F_I x \quad (S8)$$

where V1(x,y) is an exclusion zone for islands and V2(x,y) is the vortex barrier in between islands, with C1 and C2 setting the relative strengths of the two potentials. The Nshape parameter sets the shape of the vortex barrier as shown in figure S5(a), with Nshape=1 resulting in a broad potential well. We instead use the parameter Nshape=4 to get a narrower well. Setting C1=150, C2=2/250, and sigma=0.2; the potential can be seen in figure S5 (b) with the path the vortices move shown in white. The vortices follow a path from the center of an island triangle and cross the barrier through the center of the edge of a triangle, demonstrating that they follow the intended path.

Euler's method is once again used, in the overdamped limit, only broken up into x and y components.

$$\dot{x}_i(t) = \frac{1}{\eta}\left(-\frac{\partial V(x_i(t), y_i(t))}{\partial x_i} + \epsilon_{xi}(t) + B\sum_{j=1}^{N}\frac{(x_i(t) - x_j(t))}{r_{ij}(t)}U_0(r_{ij}(t))\right). \quad (S9)$$

$$\dot{y}_i(t) = \frac{1}{\eta}\left(-\frac{\partial V(x_i(t), y_i(t))}{\partial y_i} + \epsilon_{yi}(t) + B\sum_{j=1}^{N}\frac{(y_i(t) - y_j(t))}{r_{ij}(t)}U_0(r_{ij}(t))\right). \quad (S10)$$

It is then solved by performing the following operations repeatedly with periodic boundary conditions

$$x_i(t+\Delta t) = x_i(t) + \dot{x}_i(t)\Delta t. \quad (S11)$$

$$y_i(t+\Delta t) = y_i(t) + \dot{y}_i(t)\Delta t. \quad (S12)$$

Where $r_{ij}$ is the distance between vortices $i$ and $j$. N vortices are randomly seeded into 240 wells and then slowly annealed into a low energy configuration. The results when B/A=8 are shown in figure 4 in the main text, with the intermediate motion type also shown. The lattice flow motion at f=0.20 are shown in figure S5. Fig. S5 and Fig. 4(b) were both taken 10,000 time steps after the current was set, to allow structure changes due to the driving current to occur. The time between the two snapshots is approximately the time necessary for a vortex to cross from one pinning site to another, given by $\Delta t = 1.5 t_I$ in Fig. 4(b) and $\Delta t = 1.0\ t_I$ in Fig. S6, where $t_I = \frac{2\eta}{3F_I}$.

Acknowledgments

This work was supported by the DOE BES under DE-SC001249 and was carried out in part in the Frederick Seitz Materials Research Laboratory, University of Illinois at Urbana-Champaign. IMS acknowledges support from the Sloan Foundation. We would like to thank C. J. Olson Reichhardt and C. Reichhardt for their correspondence.

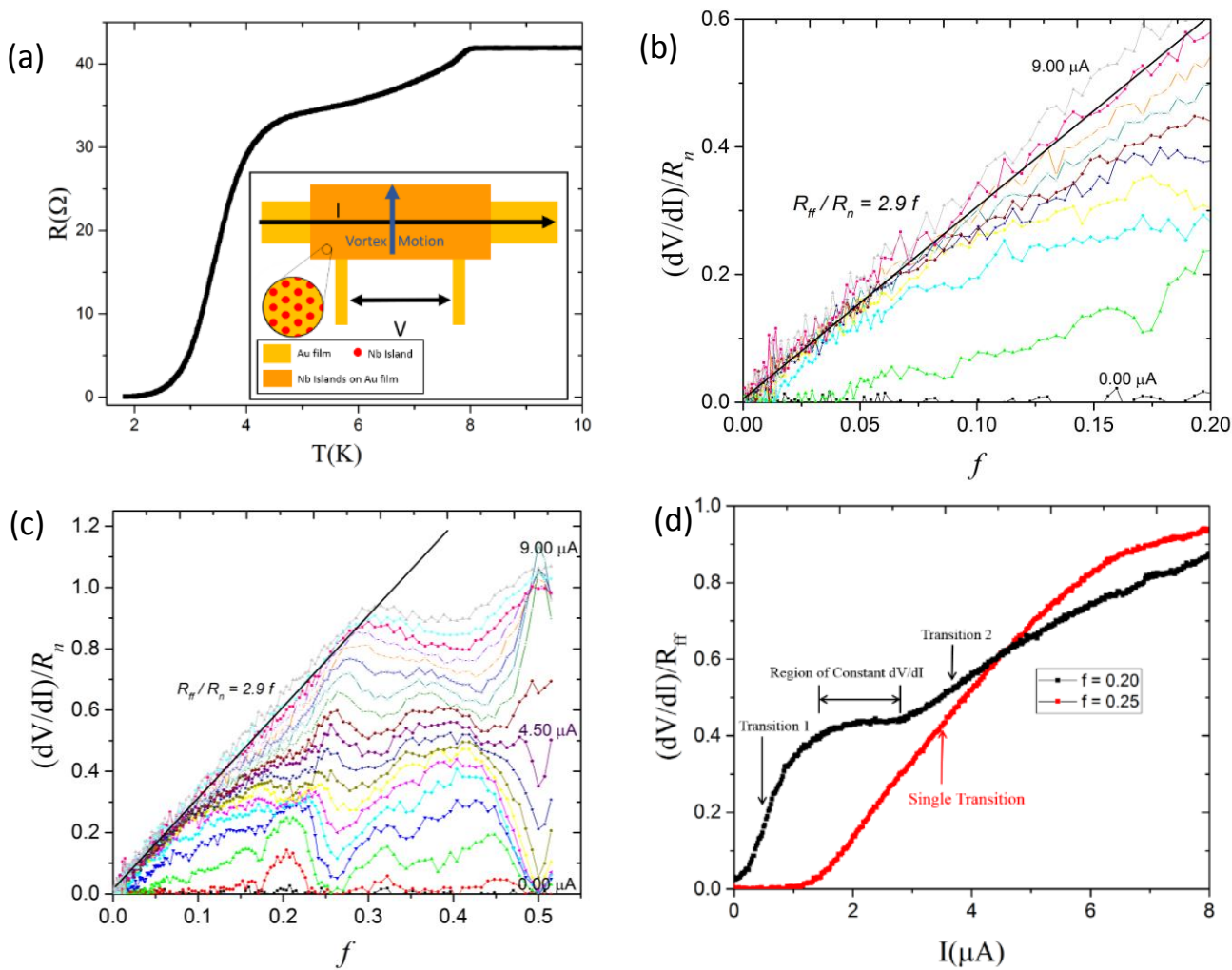

**FIG. 1. I-V measurements** (a) Resistance vs temperature for a 240 nm edge-to-edge spaced array, showing the superconducting transition. Inset shows the four point measurement configuration. (b) Normalized differential resistance $dV/dI$ at $0.5$ $\mu A$ current intervals as a function magnetic filling f. As current increases at low fillings, there is a rapid transition from $dV/dI=0$ to flux flow resistance(black line, $R_{ff}/R_N$). (c) $dV/dI$ vs $f$ at $0.5$ $\mu A$ current intervals, for a larger range than (b). At higher fillings, dips associated with commensurate fillings are visible at $f=1/6$, $1/4$, $1/3$, and $1/2$. Intermediate clusterings of lines are visible at $f=0.20$ and $f=0.35$. (d) $dV/dI$ vs $I$. The black curve for $f=0.20$ undergoes a two-step transition with an intermediate region of constant $dV/dI$ in between. The red curve for $f=0.20$ undergoes a single transition.

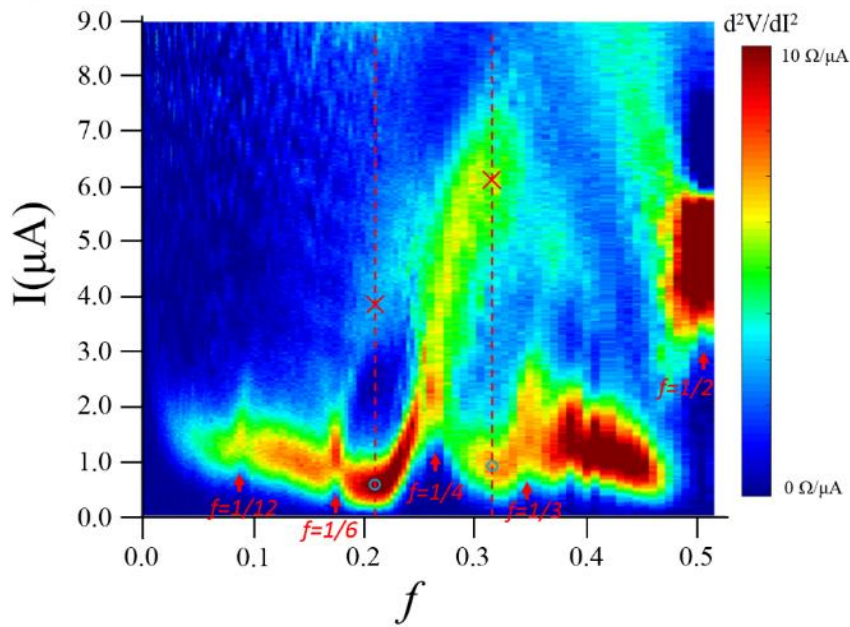

**FIG. 2. d²V/dI² as a function of current and frustration.**
Transition steps can be mapped using $d^2V/dI^2$, with higher values (brighter colors) corresponding to a transition step. Commensurate fillings are marked with red arrows and undergo a single transition. Examples of incommensurate fillings are marked with a red dashed line and undergo two transitions, the first marked with a blue circle and the second marked with a red X.

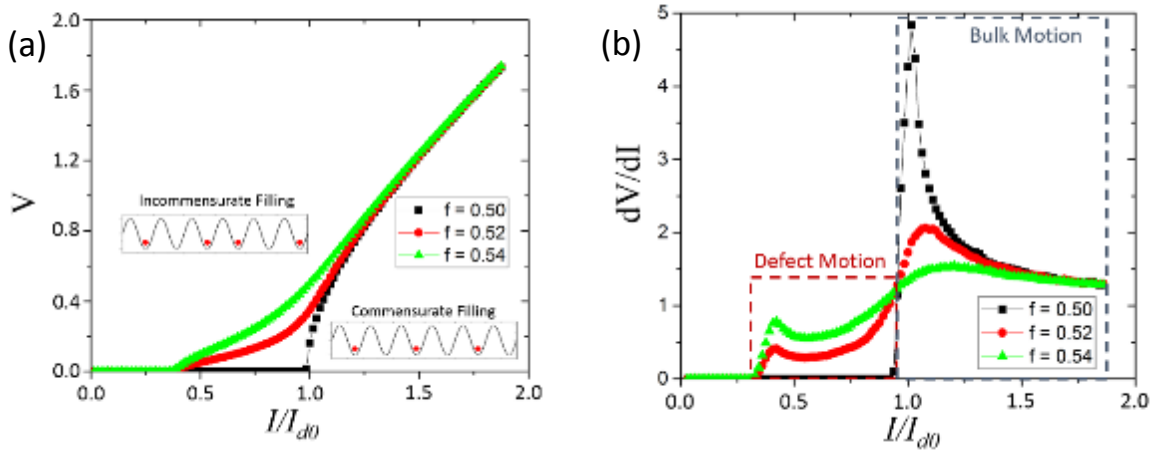

**FIG. 3. One dimensional simulation of array** (a) Simulated *I-V* behavior for commensurate ($f=0.5$) and incommensurate fillings ($f=0.52, 0.54$). Right inset shows the half filling arrangement. Left inset shows an incommensurate filling with defects in the form of vortices in adjacent wells. (b) Simulated $dV/dI$ measurements. At commensurate values, the system rapidly transitions from pinned to bulk vortex motion. As field is increased ($f=0.52, 0.54$), defects are added. This results in a transition from pinned to defect motion to bulk vortex motion. . $I_{d0}$ is the de-pinning current in the single particle limit.

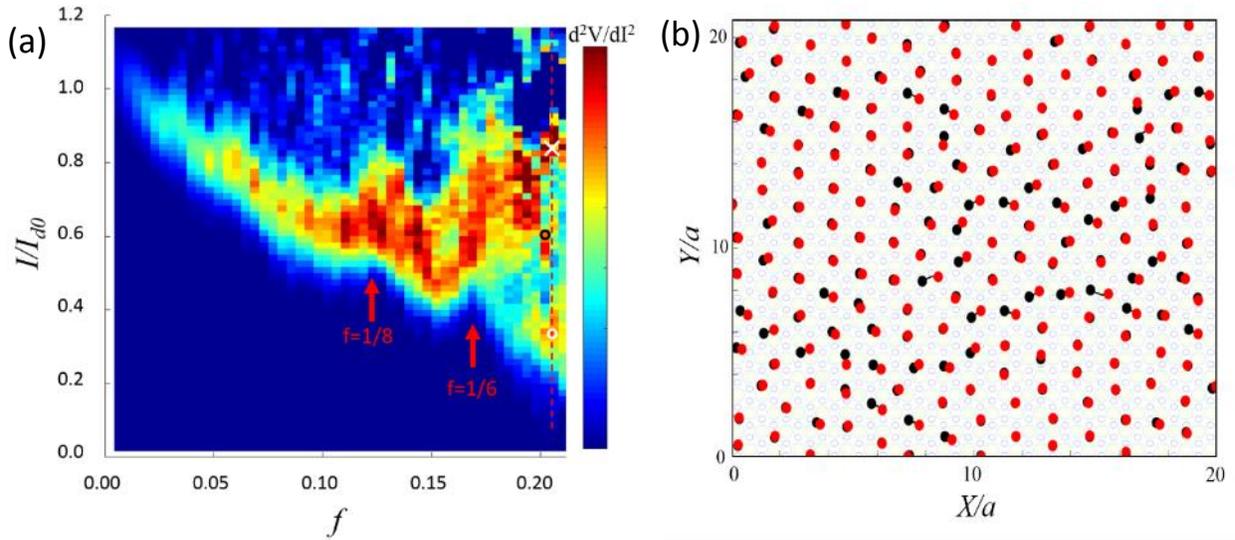

**FIG. 4. 2D simulation of triangular array.** (a) Simulated $d^2V/dI^2$. At $f=1/6$, a second transition splits off from the de-pinning current when filling is increased. An incommensurate filling is shown with a dashed line. A white circle shows the first transition. A white X shows the second transition. $I_{d0}$ is the de-pinning current in the single particle limit. (b) Time evolution of vortex motion at $f=0.20$ with an applied current of $I=0.6$, for data shown as a black circle in (a) Black circles are simulated vortices and red circles show their position a short time afterwards (with a black line showing the path in between). This simulation is in the defect motion regime, with most defect motion occurring on the interface between 2 different crystalline structures.

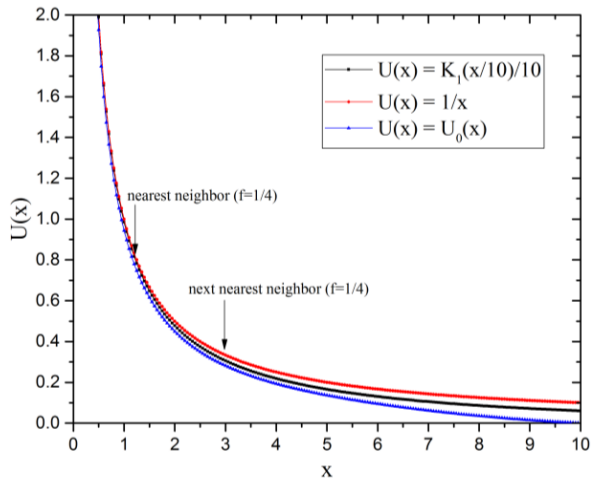

**FIG. S1. Vortex Vortex Repulsion as a function of distance** The different functions discussed as repulsion terms are shown. The function $U_0(x)$ discussed in equation S1 is a decent approximation of the predicted repulsion terms for nearest neighbor and next nearest neighbor interactions.

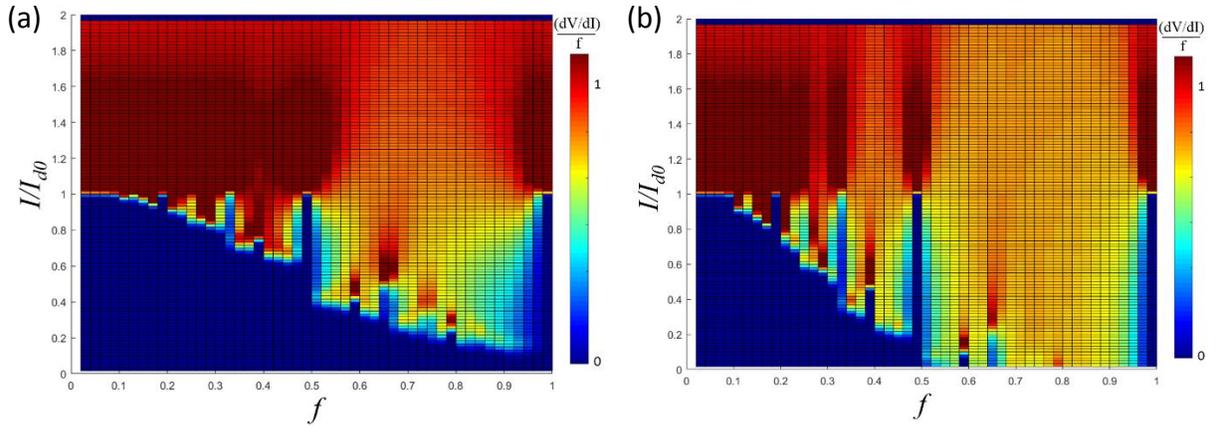

**FIG. S2. Simulated 1D dV/dI as a function of current and field.** (a) B/A=2.0, which is used in Figure 3, has visible intermediate steps associated with defect motion near commensurate fillings, most prominently just above *f=0.5* and just below *f=1.0*. (b) B/A=6.0 has a much stiffer lattice, resulting in narrower regions with two steps. Instead, the stiffer lattice favors lattice motion.

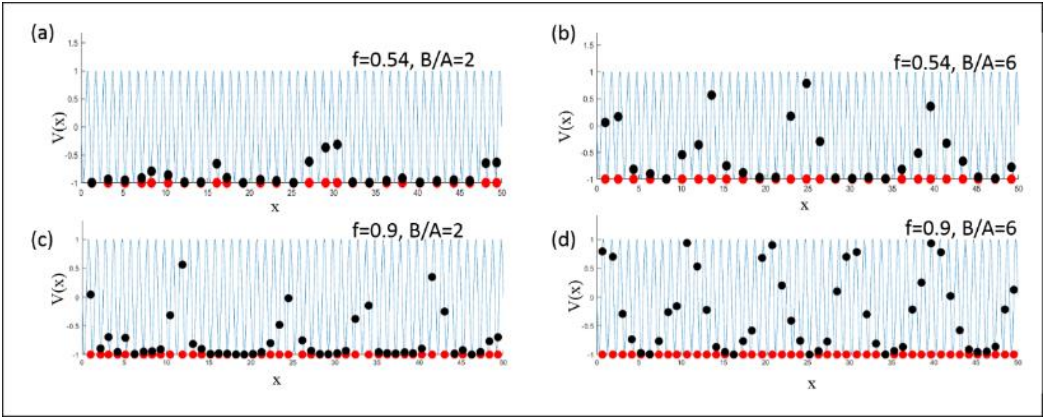

**FIG S3. Simulated 1D vortex arrangement as a function of stiffness and repulsion.** Blue lines represent the periodic potential, the red dots at the bottom of each graph show the x position of vortices, and the black dots show the energy of the vortices as well as x position. (a) f=0.54 and B/A=2.0 yields defects in the form of a pair of vortices in adjacent wells. These defects separate ordered regions with half the wells occupied. (b) f=0.54 and B/A=6.0 vortices do not sit in the wells and are not separated by integer well periods. Disorder appears to manifest in a quasiperiodic structure rather than in the interface between two ordered domains as in (a). (c) has defects appear as empty wells separating domains with every well filled. (d) Greater vortex-vortex repulsion once again yields a quasiperiodic structure. It is notable that (a) and (c) exhibit a two step transition in figure S2 (a), but (b) and (d) undergo a single transition in figure S2(b).

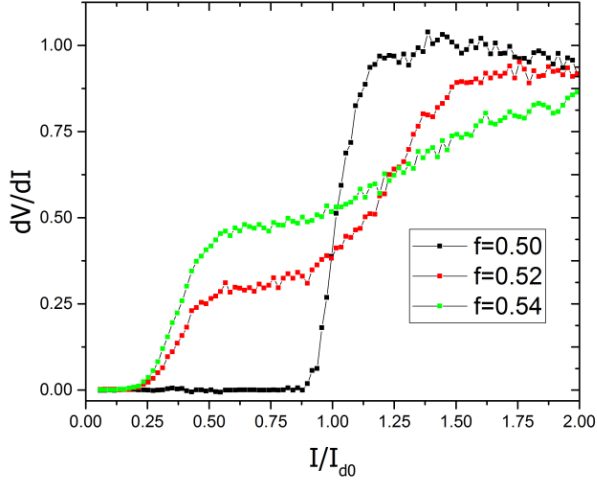

**FIG S4. Simulated 1D dV/dI with history dependent dissipative force.** The array undergoes a single transition from pinned to lattice flow for commensurate fillings. It undergoes a two step transition from pinned to defect motion to lattice flow for incommensurate fillings. The inclusion of a history dependent dissipative force removes the differential resistance peak at each step.

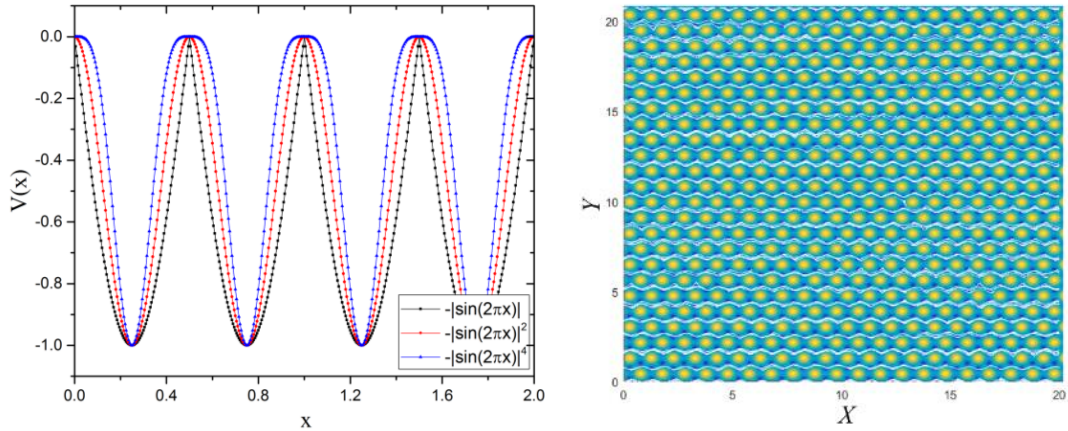

**FIG. S5. 2D Potential Well.** (a) The effect of $N_{shape}$ on periodic potential. Higher values of $N_{shape}$ yield narrower wells for vortices to rest in. (b) The resulting periodic potential is shown with wells in blue and islands in yellow. The path of vortices in a simulated lattice is shown in white. The exclusion potential around the islands results in vortices moving between the centers of the adjacent wells, which requires overcoming the potential provided by equation S7.

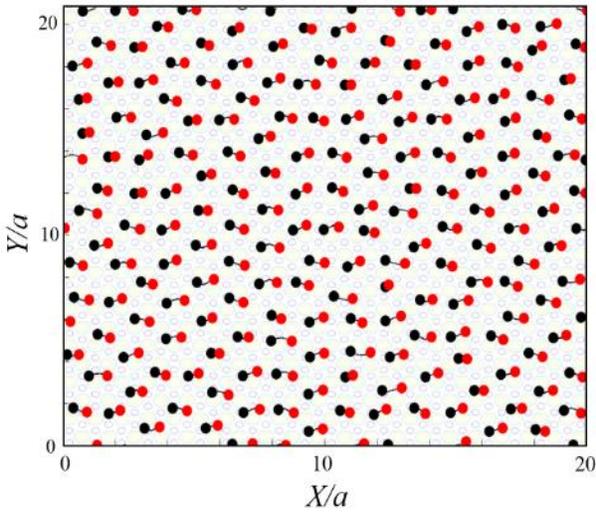

**FIG. S6. 2D simulated Flux Flow.** Flux Flow is simulated using a value of *I=2.2* and *f=0.20*. Unlike the defect motion regime shown in Figure 4 (b) with I=0.6, all vortices are in motion at once.